\documentclass[smallcondensed]{svjour3}

\smartqed

\usepackage{amsmath,amssymb,graphicx}

\newcommand{\eftnopi}{\mbox{EFT($\not \! \pi$)}}
\newcommand{\ND}{N^\dagger}
\newcommand{\CSing}{{\cal C}_0^{(^1 \! S_0)}}
\newcommand{\CTrip}{{\cal C}_0^{(^3 \! S_1)}}
\newcommand{\PSing}{P_a^{(^1 \! S_0)}}
\newcommand{\PTrip}{P_i^{(^3 \! S_1)}}
\newcommand{\aSing}{a^{(^1 \! S_0)}}
\newcommand{\aTrip}{a^{(^3 \! S_1)}}

\newcommand{\VS}{\vec{\sigma}}
\newcommand{\VT}{\vec{\tau}}
\newcommand{\LRD}{\stackrel{\leftrightarrow}{D}}
\newcommand{\aR}{{\cal C}^{(^3 \! S_1-^1 \! P_1)}}
\newcommand{\bR}{{\cal C}^{(^1 \! S_0-^3 \! P_0)}_{(\Delta I=0)}}
\newcommand{\cR}{{\cal C}^{(^1 \! S_0-^3 \! P_0)}_{(\Delta I=1)}}
\newcommand{\dR}{{\cal C}^{(^1 \! S_0-^3 \! P_0)}_{(\Delta I=2)}}
\newcommand{\eR}{{\cal C}^{(^3 \! S_1-^3 \! P_1)}}

\newcommand{\bs}[1]{\boldsymbol{#1}}

\begin{document}

\title{Hadronic parity violation in pionless effective field theory\thanks{Based on work in collaboration with D.~R.~Phillips and R.~P.~Springer \cite{Phillips:2008hn,Schindler:2009wd}.}}

\author{Matthias~R.~Schindler}

\institute{Matthias~R.~Schindler \at
              Department of Physics and Astronomy \\
              Ohio University\\
	      Athens, OH 45701\\
              \emph{Present address:} Department of Physics\\
	      The George Washington University\\
	      Washington, DC 20052 \\
	      Tel.: +1-202-994-3864\\
              Fax: +1-202-994-3001\\
              \email{schindle@gwu.edu}  \\
}

\date{}

\maketitle

\begin{abstract}

We present results for two-body observables that are sensitive to the parity-violating component of nucleon-nucleon interactions. These interactions are studied using an effective field theory in which the only dynamic degrees of freedom are nucleon fields. The observables we study are cross-section asymmetries in nucleon-nucleon scattering and asymmetries and induced polarizations in the process $np\to d\gamma$.

\end{abstract}

\section{Introduction}
\label{Sec:Intro}

The low-energy interaction between hadrons contains a parity-violating (PV) component due to the weak interactions of quarks contained in the hadrons. Hadronic parity violation has received renewed interest with a number of ongoing and planned experiments \cite{JLAB,HIGS,Markoff:2005dm,Stiliaris:2005hh,Lauss:2006es,Bass:2009fs} as well as  theoretical developments. Traditionally, there are two different theoretical approaches to hadronic parity violation. The most commonly used framework is the one-meson exchange model of Desplanques, Donoghue and Holstein (DDH) \cite{Desplanques:1979hn}. In a different approach by Danilov \cite{Danilov} the PV hadronic interactions are described in terms of 5 low-energy amplitudes. More recently, effective field theories (EFTs) have been applied to the problem of hadronic parity violation \cite{Kaplan:1992vj,Kaplan:1998xi,Savage:1998rx,Savage:2000iv,Zhu:2004vw}. Here we present results of calculations in the framework of pionless effective field theory (\eftnopi). In {\eftnopi} only nucleon fields appear as dynamic degrees of freedom. At the energies of many of the current and planned experiments the pion cannot be resolved and can be integrated out. The remaining interactions are given in terms of 5 $NN$ contact terms, each accompanied by an unknown low-energy constand (LEC). While the LECs cannot be predicted within the theory, once they have been determined by comparison with a sufficient number of experimental data the theory has predictive power. As an EFT this approach is model-independent and allows for a systematic improvement of the results. The results presented here constitute an important step in the determination of the PV low-energy constants.

\section{Lagrangians}
\label{Sec:Lags}

In {\eftnopi} all interactions are described by contact terms. In the parity-conserving sector the Lagrangian is given by
\begin{align}\label{Lag:PC}
\mathcal{L}_{PC} & =  \ND(i D_0 + \frac{\vec{D}^2}{2M})N +\frac{e}{2M}\ND (\kappa_0+\tau_3\kappa_1)\,\bs{\sigma}\cdot\bs{B} N \notag\\
&-\CSing (N^T \PSing N)^\dagger (N^T \PSing N) -\CTrip (N^T \PTrip N)^\dagger (N^T \PTrip N ) + \ldots,
\end{align}
where $\sigma_i$ and $\tau_a$ are SU(2) Pauli matrices in spin and isospin space, respectively, $D_\mu N$ is the nucleon covariant derivative,
\begin{equation}
D_\mu N = \partial_\mu N +ie\frac{1+\tau_3}{2}A_\mu N,
\end{equation}
and $\kappa_0$ and $\kappa_1$ are the isoscalar and isovector nucleon
magnetic moments. The partial wave projection operators are defined through \cite{Kaplan:1998sz}
\begin{equation}
\PSing=\frac{1}{\sqrt{8}} \tau_2 \tau_a \sigma_2 \ , \quad 
\PTrip=\frac{1}{\sqrt{8}} \tau_2 \sigma_2 \sigma_i  \ .
\end{equation}
In the power-divergence subtraction scheme \cite{Kaplan:1998tg,Kaplan:1998we}, which we employ in our calculations, the low-energy couplings are given by 
\begin{align}
\CSing &= \frac{4\pi}{M}\frac{1}{\frac{1}{\aSing}-\mu}\ , \label{Lag:CSing}\\
\CTrip &= \frac{4\pi}{M}\frac{1}{\frac{1}{\aTrip}-\mu}\ , \label{Lag:CTrip}
\end{align}
with $\aSing$ ($\aTrip$) the $^1S_0$ ($^3S_1$) scattering length and $\mu$ the renormalization point.

For the leading-order PV Lagrangian we use the formulation given in \cite{Phillips:2008hn},
\begin{align}\label{Lag:PV}
\mathcal{L}_{PV}=  -  & \left[ \aR \left(N^T\sigma_2 \ \VS \tau_2 N \right)^\dagger 
\cdot  \left(N^T \sigma_2  \tau_2 i\LRD N\right) \right. \notag\\
& +\bR \left(N^T\sigma_2 \tau_2 \VT N\right)^\dagger  
\left(N^T\sigma_2 \ \VS \cdot \tau_2 \VT i\LRD  N\right) \notag\\
& +\cR \ \epsilon^{3ab} \left(N^T\sigma_2 \tau_2 \tau^a N\right)^\dagger 
\left(N^T \sigma_2  \ \VS\cdot \tau_2 \tau^b \LRD N\right) \notag\\
& +\dR \ \mathcal{I}^{ab} \left(N^T\sigma_2 \tau_2 \tau^a N\right)^\dagger 
\left(N^T \sigma_2 \ \VS\cdot \tau_2 \tau^b i \LRD N\right) \notag\\
& +\left. \eR \ \epsilon^{ijk} \left(N^T\sigma_2 \sigma^i \tau_2 N\right)^\dagger 
\left(N^T \sigma_2 \sigma^k \tau_2 \tau_3 \LRD{}^{\!j} N\right) \right] + h.c.,
\end{align}
where $a\, \mathcal{O}\LRD b = a\,\mathcal{O}\vec D b - (\vec D a)\mathcal{O} b$ with $\mathcal{O}$ some spin-isospin-operator, and 
$$ \mathcal{I}=
\begin{pmatrix} 
1 & 0 & 0 \\
0 & 1 & 0\\
0 & 0 & -2
\end{pmatrix}. $$

\section{NN scattering}
\label{Sec:NN}

Nucleon-nucleon scattering is theoretically the most basic process in which to study hadronic parity violation. The longitudinal asymmetry
\begin{equation}\label{NN:Asym}
 A_L=\frac{\sigma_+ -\sigma_-}{\sigma_+ +\sigma_-},
\end{equation}
where $\sigma_\pm$ is the total scattering cross section of a polarized nucleon with helicity $\pm$ on an unpolarized nucleon target, would vanish if parity were conserved. With the Lagrangians given in Sec.~\ref{Sec:Lags} and neglecting Coulomb interactions for the $pp$ case we obtain at leading order
\begin{align}\label{NN:ppnn}
 A_L^{pp/nn} &=8p \frac{\mathcal{A}_{pp/nn}}{\CSing}
\end{align}
and 
\begin{align}
A_L^{np} &= 8 p \left( \frac{ \frac{d\sigma^{^1\!S_0}}{d\Omega}}{\frac{d\sigma^{^1\!S_0}}{d\Omega}+3\frac{d\sigma^{^3\!S_1}}{d\Omega}}\,\frac{{\cal A}_{np}^{^1\!S_0}}{{\cal C}_0^{^1\!S_0}} + \frac{ \frac{d\sigma^{^3\!S_1}}{d\Omega}}{\frac{d\sigma^{^1\!S_0}}{d\Omega}+3\frac{d\sigma^{^3\!S_1}}{d\Omega}} \, \frac{{\cal A}_{np}^{^3\!S_1}}{{\cal C}_0^{^3\!S_1}}\right),
\end{align}
with
\begin{align}
\mathcal{A}_{nn} &= 4\left( \bR -  \cR +  \dR \right), \label{eq:A1b}\\
\mathcal{A}_{pp} &= 4 \left( \bR + \cR + \dR \right), \label{eq:A2}\\
\mathcal{A}_{np}^{^1S_0} &= 4 \left(\bR - 2 \dR \right), \label{eq:A3}\\
\mathcal{A}_{np}^{^3S_1} &= 4 \left(\aR -2 \eR \right) \label{eq:A4}
\end{align}
linear combinations of the parameters of the Lagrangian, and $\frac{d\sigma^X}{d\Omega}$ the differential cross section in partial wave $X$. Since $A_L^{pp/nn}$ is an observable and therefore independent of the renormalization point $\mu$, the $\mu$-dependence of $\mathcal{A}_{pp/nn}$ is dictated entirely by $\CSing$. In addition, since the energy dependence of the differential cross sections in the $^1\!S_0$ and $^3\!S_1$ partial waves differ, a detailed study of the energy dependence of the $np$ asymmetry in principle allows an extraction of two different linear combinations of the PV parameters. 

The result of Eq.~(\ref{NN:ppnn}) does not take into account Coulomb effects for the $pp$ case. Coulomb interactions can be systematically incorporated in {\eftnopi}. We find that at energies for which {\eftnopi} is applicable Coulomb corrections turn out to be small and can be treated perturbatively \cite{Phillips:2008hn}.

There exist two experimental results for $A_L^{pp}$ at $13.6\,\mbox{MeV}$ \cite{Eversheim:1991tg} and $45\,\mbox{MeV}$ \cite{Kistryn:1987tq}. While the result at $13.6\,\mbox{MeV}$ can be used to extract the combination \cite{Phillips:2008hn}
\begin{equation}
 \mathcal{A}_{pp}(\mu=m_\pi)=(1.3 \pm 0.3) \times 10^{-14}\, \mbox{MeV}^{-3},
\end{equation}
the energy of $45\,\mbox{MeV}$ lies outside the domain of applicability of \eftnopi.

\section{Radiative neutron capture}
\label{Sec:npdgamma}

By choosing suitable neutron polarizations the reaction $np\to d\gamma$ allows access to two different PV observables. For polarized neutrons one can determine the photon asymmetry $A_\gamma$ at threshold, which is defined through
\begin{equation}
 \frac{1}{\Gamma}\frac{d\Gamma}{d \theta}= 1+A_\gamma \cos \theta,
\end{equation}
where $\Gamma$ is the $np\to d\gamma$ width and $\theta$ the angle between the neutron polarization and the direction of the outgoing photon momentum. At leading order we obtain
\begin{equation}
 A_\gamma=\frac{32}{3}\,\frac{M}{\kappa_1\left(1-\gamma \aSing\right)}\,\frac{\eR}{\CTrip}\ .
\end{equation}
Note again the appearance of a ratio of PV and PC couplings, indicating that any renormalization point dependence in the weak couplings is driven by those in the strong coupling. We would also like to point out that we do not require phenomenological wave functions for the deuteron, but treat the deuteron field consistently in the EFT framework. There is an ongoing experimental effort at the Spallation Neutron Source to measure $A_\gamma$ to a sensitivity of $\delta A_\gamma=1\times 10^{-8}$ (see e.g.~\cite{Crawford}).

For unpolarized neutron capture the PV interactions induce a circular polarization of the outgoing photon $P_\gamma$,
\begin{equation}
 P_\gamma = \frac{\sigma_+-\sigma_-}{\sigma_+ + \sigma_-},
\end{equation}
where $\sigma_{+/-}$ is the total cross section for photons with positive/negative helicity. The leading-order result is given by
\begin{align}\label{Res:PolRes}
P_\gamma= & -16\frac{M}{\kappa_1\left(1-\gamma\aSing\right)}\,\left[ \left(1-\frac{2}{3}\gamma\aSing\right)\frac{\aR}{\CTrip} \right. \notag\\
& \left. +\frac{\gamma\aSing}{3}\frac{\bR-2\dR}{\CSing} \right].
\end{align}
The asymmetry in the inverse reaction might be experimentally more feasible and is equal to $P_\gamma$ for exactly reversed kinematics. Current experimental results are consistent with zero in both cases \cite{Alberi:1988fd,Knyaz'kov:1984zz,Earle:1988fc}.

\section{Conclusions}
\label{Sec:Conc}

We have presented results for two-body PV observables in the framework of pionless effective field theory,\footnote{We have also calculated the neutron capture observables in a formulation of {\eftnopi} that includes explicit dibaryon fields \cite{Schindler:2009wd}.} in particular asymmetries in $NN$ scattering and polarized radiative neutron capture, and an induced photon polarization in unpolarized neutron capture. These results provide an important step in the determination of the unknown low-energy constants in the PV Lagrangian of Eq.~(\ref{Lag:PV}). In order to determine all five parameters it is necessary to also consider few-body processes. In the future we will look at observables such as, e.g., asymmetries in $n d\to t\gamma$ or $n \alpha$ spin rotation for which experiments are ongoing or being considered.

\begin{acknowledgements}
I thank D.R.~Phillips and R.~P.~Springer for their collaboration on the topics presented here and many interesting discussions, as well as useful comments on the manuscript. This work was supported by the US Department of Energy under grant DE-FG02-93ER40756 and grant DE-FG02-05ER41368. I also acknowledge financial support by the National Science Foundation (CAREER grant PHY-0645498) and US Department of Energy (DE-FG02-95ER-40907).
\end{acknowledgements}


\begin{thebibliography}{}

\bibitem{Phillips:2008hn}
  D.~R.~Phillips, M.~R.~Schindler and R.~P.~Springer,
  Nucl.\ Phys.\  A {\bf 822}, 1 (2009)
  [arXiv:0812.2073 [nucl-th]].

\bibitem{Schindler:2009wd}
  M.~R.~Schindler and R.~P.~Springer,
  arXiv:0907.5358 [nucl-th].



\bibitem{JLAB}
Ch.~Sinclair et al.,
``Letter-of-Intent 00-002 for PAC 17: Study of the Parity Nonconserving
Force Between Nucleons Through Deutron Photodisintegration''.

\bibitem{HIGS}
B.~Wojtsekhowski and W.T.H.~van Oers,
``Summary of the Working Group Meeting on Parity Violation in Deuteron
Photodisintegration with Circularly Polarized Photons,''
13-14 April, 2000, Jefferson Lab.

\bibitem{Markoff:2005dm}
D.M.~Markoff,
J.\ Res. \ Natl \ Inst.\ Stan.\ Tech.\ {\bf 110}, 209 (2005).

\bibitem{Stiliaris:2005hh}
  E.~Stiliaris,
  Eur.\ Phys.\ J.\  A {\bf 24S2}, 175 (2005).

\bibitem{Lauss:2006es}
 B.~Lauss {\it et al.},
 AIP Conf.\ Proc.\  {\bf 842}, 790 (2006)
 [arXiv:nucl-ex/0601004].

\bibitem{Bass:2009fs}
  C.~D.~Bass {\it et al.},
  arXiv:0905.0395 [nucl-ex].

\bibitem{Desplanques:1979hn}
 B.~Desplanques, J.~F.~Donoghue and B.~R.~Holstein,
 Annals Phys.\  {\bf 124}, 449 (1980).

\bibitem{Danilov}
 G.~S.~Danilov, Phys. Lett. {\bf 18}, 40 (1965); 
  Phys.\ Lett.\  {\bf B35}, 579 (1971).
 Sov.\ J.\ Nucl.\ Phys.  {\bf 14}, 443 (1972). 


\bibitem{Kaplan:1992vj}
  D.~B.~Kaplan and M.~J.~Savage,
  Nucl.\ Phys.\  A {\bf 556}, 653 (1993)
  [Erratum-ibid.\  A {\bf 570}, 833 (1994\ ERRAT,A580,679.1994)].



\bibitem{Kaplan:1998xi}
  D.~B.~Kaplan, M.~J.~Savage, R.~P.~Springer and M.~B.~Wise,
  Phys.\ Lett.\  B {\bf 449}, 1 (1999)
  [arXiv:nucl-th/9807081].

\bibitem{Savage:1998rx}
  M.~J.~Savage and R.~P.~Springer,
  Nucl.\ Phys.\  A {\bf 644},  235 (1998)
  [Erratum-ibid.\  A {\bf 657},  457  (1999)]
  [arXiv:nucl-th/9807014].

\bibitem{Savage:2000iv}
  M.~J.~Savage,
  Nucl.\ Phys.\  A {\bf 695}, 365 (2001)
  [arXiv:nucl-th/0012043].

\bibitem{Zhu:2004vw}
  S.~L.~Zhu, C.~M.~Maekawa, B.~R.~Holstein, M.~J.~Ramsey-Musolf and U.~van
Kolck,
  Nucl.\ Phys.\  A {\bf 748}, 435 (2005)
  [arXiv:nucl-th/0407087].

\bibitem{Kaplan:1998sz}
  D.~B.~Kaplan, M.~J.~Savage and M.~B.~Wise,
  Phys.\ Rev.\  C {\bf 59}, 617 (1999)
  [arXiv:nucl-th/9804032].

\bibitem{Kaplan:1998tg}
  D.~B.~Kaplan, M.~J.~Savage and M.~B.~Wise,
  Phys.\ Lett.\  B {\bf 424}, 390 (1998)
  [arXiv:nucl-th/9801034].

\bibitem{Kaplan:1998we}
  D.~B.~Kaplan, M.~J.~Savage and M.~B.~Wise,
  Nucl.\ Phys.\  B {\bf 534}, 329 (1998)
  [arXiv:nucl-th/9802075].

\bibitem{Eversheim:1991tg}
 P.~D.~Eversheim {\it et al.},
 Phys.\ Lett.\  B {\bf 256} (1991) 11; 
 P.~D.~Eversheim {\it et al.},
  Spring Meeting of the DPG, Salzburg 1992, Abstract in Verhandlungen der DPG (1992) 59;
 W.~Haeberli and B.~R.~Holstein,
  arXiv:nucl-th/9510062.

\bibitem{Kistryn:1987tq}
 S.~Kistryn {\it et al.},
 Phys.\ Rev.\ Lett.\  {\bf 58}, 1616 (1987).

\bibitem{Crawford}
  C.~B.~Crawford, 4th International Workshop ``From Parity Violation to Hadronic Structure and more\ldots'', 22-26 June 2009, Bar Harbor, \verb+http://web.mit.edu/pavi09/talks/Crawford_neutron_cap_pv.pdf+.

\bibitem{Alberi:1988fd}
 J.~Alberi {\it et al.},
 Can.\ J.\ Phys.\  {\bf 66} (1988) 542.


\bibitem{Knyaz'kov:1984zz}
  V.~A.~Knyaz'kov {\it et al.},
  Nucl.\ Phys.\  A {\bf 417}, 209 (1984).

\bibitem{Earle:1988fc}
  E.~D.~Earle {\it et al.},
  Can.\ J.\ Phys.\  {\bf 66} (1988) 534.


\end{thebibliography}
\end{document}